\documentstyle[sprocl,epsf]{article}

\def\NPB{{\em Nucl. Phys.} B}
\def\PLB{{\em Phys. Lett.}  B}

\def\PRD{{\em Phys. Rev.} D}

\newcommand{\be}{\begin{equation}}
\newcommand{\ee}{\end{equation}}
\newcommand{\beq}{\begin{eqnarray}}
\newcommand{\eeq}{\end{eqnarray}}

\begin{document}

\title{The Infrared behaviour of the gluon propagator in SU(2) and SU(3)
 without lattice 
Gribov copies\footnote[1]{Talk given by C. Alexandrou}}

\vspace*{-0.8cm}

\author{C. Alexandrou$^a$, Ph. de Forcrand$^{b,c}$, E. Follana$^a$}
\address{$^a$Department of Physics, 
        University of Cyprus, CY-1678 Nicosia, Cyprus\\
$^b$Inst. f\"ur Theoretische Physik, ETH H\"onggerberg, CH-8093 Z\"urich,
Switzerland \\
$^c$ CERN, Theory Division, CH-1211 Geneva 23, Switzerland
}
 
\vspace*{-0.5cm}

\maketitle\abstracts{We present lattice results for the gluon propagator 
for SU(2) and SU(3) 
in the Laplacian gauge which avoids lattice Gribov copies. In SU(3) we 
compare with
the most recent lattice calculation  
in Landau gauge and with various approximate solutions of
the Dyson Schwinger equations (DSE).
}

\vspace*{-0.5cm}

\noindent
{\bf Introduction}

We first summarize the
results obtained within the Landau gauge\cite{summary}:
By solving approximately the DSE,
 Mandelstam found an infrared enhanced gluon propagator of the form
$
 D(q^2)  \stackrel{ q^ \rightarrow  0} \sim \frac{1}{q^4}.
$
Avoiding gauge copies, Gribov obtained
$  D(q^2) \sim \frac{q^2}{q^4 + m^4} $.
Using the ``pinch technique'', Cornwall~\cite{Cornwall}
obtained  a solution which
fulfills the Ward identities, allows a dynamical mass 
generation, and
also predicts a finite value for $D(0)\equiv D(q^2=0)$
 consistent with our data.

Early results for  the gluon propagator obtained directly
 from Lattice QCD on  small lattices ~\cite{MO}
were interpreted in terms of a massive scalar
propagator.
Results on larger lattices
were accounted for by assuming a positive anomalous
dimension~\cite{Marenzoni}: 
$
D(q^2) \sim \frac{1}{q^{2(1+\alpha)} + m^2}
$.
A recent, detailed study
of the gluon propagator uses very large lattices~\cite{Williams}.
Since we want to compare our results with these, we follow closely their
analysis and refer to Refs.~\cite{Williams,AFF} for details.
In the Laplacian gauge,
the longitudinal part of the gluon propagator does not vanish;
 the  transverse scalar function $D(q^2)$ can be  extracted from
${\cal D}^{ab}_{\mu\nu}(q)$ as  
$
D(q^2) = \frac{1}{3}\biggl\{\sum_\mu \frac{1}{8} 
                     \sum_a {\cal D}^{aa}_{\mu\mu}(q) \biggr\}
         -\frac{1}{3}  \frac{F(q^2)}{q^2} ~, 
$
where $F(q^2)$ is determined by projecting the longitudinal part of 
${\cal D}^{aa}_{\mu\nu}(q)$ using the symmetric tensor $q^\mu q^\nu$. 

\noindent
{\bf Gauge Fixing Procedure}

Previous lattice studies all 
fixed to Landau gauge by using a local iterative maximization algorithm,
which converges to any one of many local maxima
(lattice Gribov copy), but fails to determine the global 
one. 
To overcome this problem, we use a 
different gauge condition, the Laplacian gauge~\cite{VW}, 
which is Lorentz-symmetric and gives a smooth gauge field like 
the Landau gauge, but which specifies the gauge unambiguously. 
We consider the maximization of 
$ Q = {\rm Re} \sum_{x,\mu} {\rm Tr} 
\left [ g(x)U_\mu(x)g^\dagger(x+\hat{\mu}) - g(x)g(x)^\dagger \right] . 
$  
If one relaxes the requirement that $g \in SU(N)$, maximizing 
$Q$ is equivalent to minimizing the quadratic form $ 
\sum_{xy} f^*_x \Delta_{xy} f_y $, with $\Delta(U)$ the covariant 
Laplacian. 
Using the $(N-1)$ lowest-lying eigenvectors ${\bf f}_i(x)$
 of $\Delta(U)$, one can 
fix the gauge uniquely by requiring $\forall x, f_i^i(x) \in {\bf R},
f_i^j(x)=0, j=(i+1),..,N$~\cite{AFF}.

\noindent
{\bf Results}

In Fig.1 we show the transverse gluon propagator for SU(2) 
Yang-Mills theory in two different volumes;
$m_0 \equiv \sqrt{D(0)^{-1}}$ for the $16^4$ lattice. 
Changing the volume has little effect, in particular on $D(0)$.
We observe similarly small volume effects in SU(3).
This is strikingly different from Landau gauge, where
Zwanziger has argued that $D(0)$ should vanish in the infinite
lattice volume limit~\cite{Zwanziger}. 
This prediction is indeed consistent with recent lattice results
in SU(2) at finite temperature~\cite{Cucchieri}.
In contrast, in the Laplacian
gauge, we find that 
$D(0)$ is finite and independent of the volume $V$
for $V$ larger than about $1/2 {\rm fm}^4 \sim D(0)^2$.
We find $D(0)=58(2)$ in lattice units at $\beta=6.0$, i.e.
$D(0)^{-1/2}=248(5)$ MeV (using $a^{-1}=1.885$~GeV), 
corresponding to a length scale of about $0.8$~fm.

In Fig.2  we compare results for the gluon propagator 
in SU(3) quenched QCD in Laplacian and Landau gauges.
($m_0 \equiv \sqrt{D(0)^{-1}}$ in the Laplacian gauge).
Scaling is checked  on the $16^3\times 32$ lattice  
for $\beta=5.8$ and $6.0$.
Making a cylindrical cut in the momenta~\cite{Williams} to minimize lattice artifacts, we find
that scaling is very well satisfied for the Laplacian gauge,
with both sets of data falling on  a universal curve~\cite{AFF}.

\begin{figure}[bp]
\begin{minipage}{6.2cm} 
\epsfxsize=5.8truecm
\epsfysize=4.8truecm
\vspace*{-0.5cm}
\mbox{\epsfbox{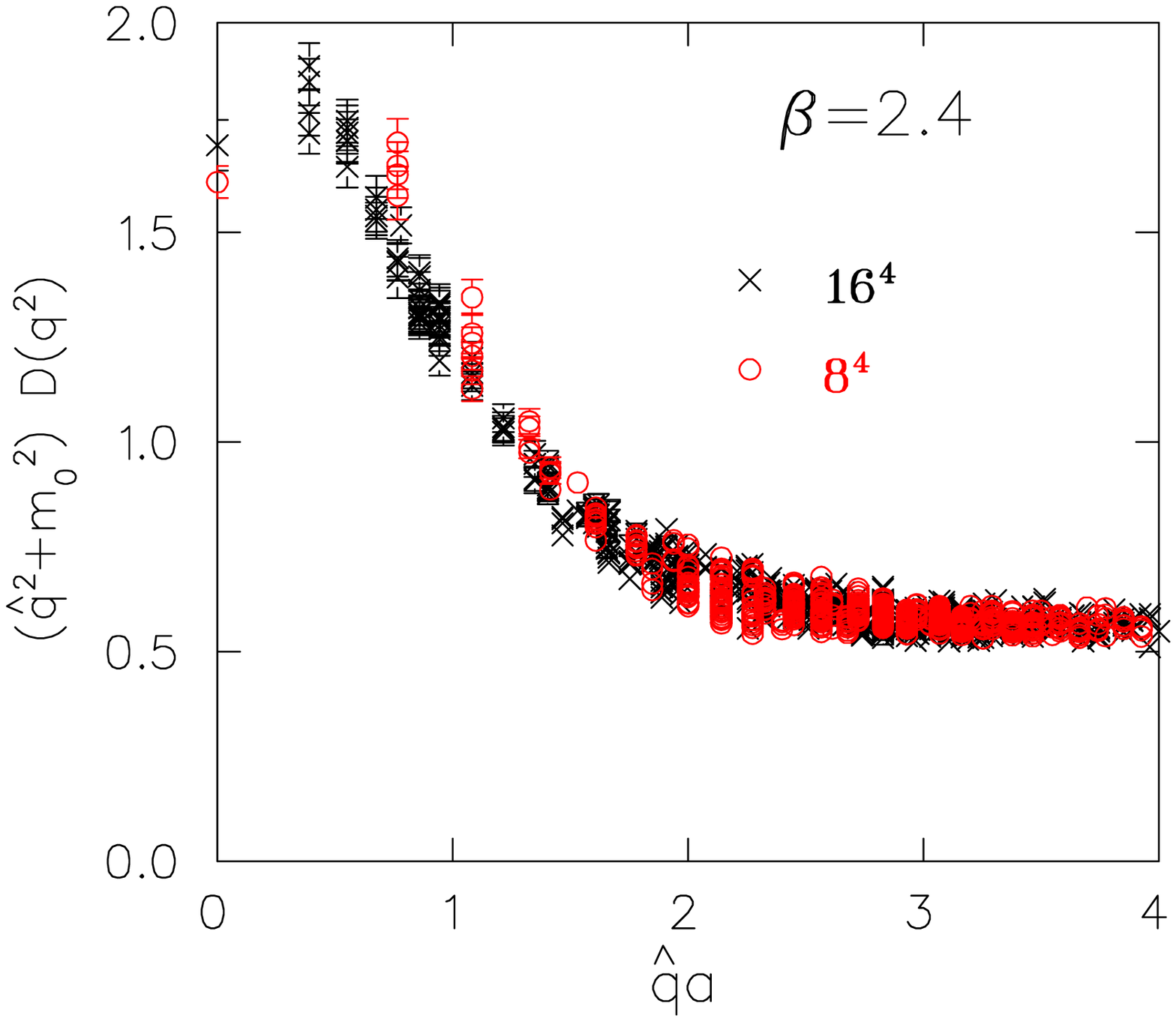}}
\vspace*{-0.5cm}
\caption{The SU(2) gluon propagator in \newline two different volumes.}
\label{Fig:su(2)}
\end{minipage} 
\begin{minipage}{5.5cm}
\epsfxsize=5.4truecm
\epsfysize=4.8truecm
\vspace*{-0.5cm}
\mbox{\epsfbox{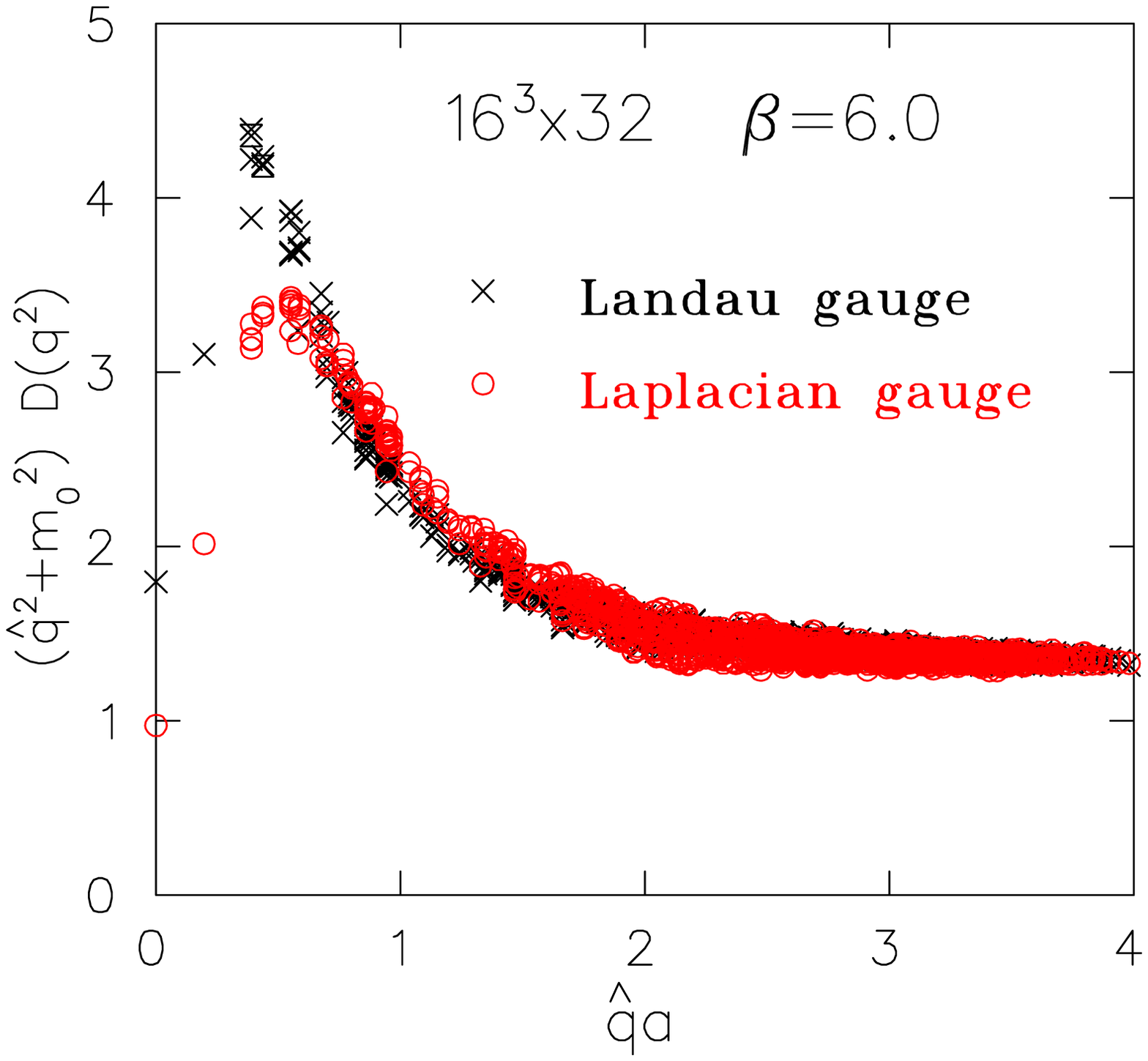}}
\vspace*{-0.5cm}
\caption{The SU(3) gluon propagator in Laplacian and Landau gauges.}
\label{Fig:compare}
\end{minipage}
\end{figure}


\small
\begin{table}[thbp]
\vspace*{-0.9cm}
\begin{center}
\begin{tabular}{|c||cccc|c|c|}
\hline
Model &Z & m& $\lambda$ or $\alpha$ & A & D(0) & $\chi^2/$d.o.f\\
\hline
Gribov & 2.63(2) & 0.203(7) & & &  0 & 5.7\\ 
Stingl &2.63(2)  &0.203(13) & 0.002 (1.100) & & 0 & 5.7 \\
Marenzoni & 2.47(3) & 0.199(6) &  0.237(5)& & 62 & 4 \\
Cornwall & 7.08(9) & 0.281(4) & 0.265(8) & & 59 & 2.5 \\
Model A &1.96(1) &0.654(17)& 2.181(67) & 8.91(41) & 43 &1.2\\ 
\hline
\end{tabular}
\end{center}
Table 1: best fit of parameter values 
to our $\beta=6.0$ data on the $16^3\times 32$ lattice.
\label{Table:fits}
\vspace*{-0.5cm}
\end{table}
\normalsize
We fit to our data the same models as considered 
 by Leinweber {\it et  al.}~\cite{Williams} in Landau gauge.
Since we have observed scaling, we use our results at the finer lattice 
spacing ($\beta=6.0$) for the fits. 
Table~1 and Fig.~3 summarize the results of the fits
to the various models.
We find that Gribov--type models are
excluded, whereas Cornwall's model is clearly favored among
all analytically motivated models.
Model ``A''~\cite{Williams}, which gives a better fit, 
is phenomenological, contains one more parameter,
and misses $D(0)$ by 25\%.
One can then use the fit to Cornwall's model to analytically continue to negative
$q^2$ and determine the gluon pole mass. This is carried out in Ref.~\cite{AFF}.

\begin{figure}[h]
\vspace*{-0.1cm}
\begin{minipage}{5.5cm}
\parbox{5.5cm}{~~~~In conclusion, we see significant modifications from Landau gauge
in the infrared.
 In particular, we
find that $D(0)$ obeys scaling, is finite, and volume independent 
for large enough volumes. 
We find support for  Cornwall's model 
which fits the momentum dependence of the propagator
rather well, whereas models with infrared enhancement of the type $1/(q^2)^2$ 
or Gribov--type suppression  are excluded.}
\end{minipage} \hfill
\begin{minipage}{5.5cm} 
\epsfxsize=6.0truecm
\epsfysize=5.truecm
\mbox{\epsfbox{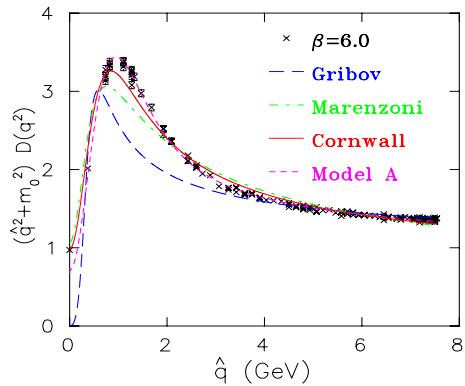}}
\vspace*{-0.8cm}
\caption{Fits to various models}
\end{minipage}
\vspace*{-0.5cm}
\end{figure}


\vspace*{-0.3cm}
\section*{References}
\vspace*{-0.2cm}
\begin{thebibliography}{99}
\bibitem{summary}J.~E.~Mandula, Phys.~Rep.~315 (1999) 273.
\bibitem{Cornwall}J.~M.~Cornwall, \PRD {\bf 26} (1982) 1453.
\bibitem{VW} J.~C.~Vink and U.~Wiese, \PLB{ \bf 289} (1992) 122.
\bibitem{MO} J.~E.~Mandula and M.~Ogilvie, \PLB {\bf 185}
  (1987) 127;R.~Gupta {\it et al.}, \PRD {\bf 36} (1987) 2813.
\bibitem{Marenzoni} P.~Marenzoni, G.~Martinelli and N.~Stella,
 \NPB {\bf 455} (1995) 339.
\bibitem{Williams} D.~Leinweber {\it et al.}, \PRD {\bf 60} (1998) 
094507; T.~G.~Williams, this volume.
\bibitem{AFF} C.~Alexandrou, Ph.~de Forcrand and E.~Follana, hep-lat/0008012.
\bibitem{Zwanziger} D.~Zwanziger, \PLB {\bf 257} (1991) 168.
\bibitem{Cucchieri}A.~Cucchieri, hep-lat/9908050.
\end{thebibliography}
\end{document}